# Rate equation analysis on the dynamics of first-order exciton Mott transition

Fumiya Sekiguchi and Ryo Shimano

*Department of Physics, The University of Tokyo, Bunkyo, Tokyo 113-0033, Japan*
*Cryogenic Research Center, The University of Tokyo, Bunkyo, Tokyo 113-0032, Japan*

We performed a rate equation analysis on the dynamics of exciton Mott transition (EMT) with assuming a detailed balance between excitons and unbound electron-hole (e-h) pairs. Based on the Saha equation with taking into account the empirical expression for the band-gap renormalization effect caused by the unbound e-h pairs, we show that the ionization ratio of excitons exhibits a bistability as a function of total e-h pair density at low temperatures. We demonstrate that an incubation time emerges in the dynamics of EMT from oversaturated exciton gas phase on the verge of the bistable region. The incubation time shows a slowing down behavior when the pair density approaches toward the saddle-node bifurcation of the hysteresis curve of the exciton ionization ratio.

Photocontrol of condensed matter systems has gained continuing interests over decades,[1-3] e.g. in semiconductors,[4,5] Mott insulators,[3,6-9] spin crossover complexes,[10-12] and superconductors.[13,14] Among diverse material systems, photoexcited electron-hole (e-h) systems in semiconductors offer a unique arena to study the rich variety of phases emergent in many particle systems, and their non-equilibrium dynamics.[4,15] One of the intriguing aspects of e-h systems is that the strength of inter-particle Coulomb interaction can be effectively controlled by changing the density of photoexcited carriers through the screening effect, by simply changing the excitation light intensity. The change of the Coulomb interaction causes a phase transition, or crossover, from the insulating exciton gas phase in the low density regime to the metallic e-h plasma in the high density regime, referred to as exciton Mott transition (EMT). There is a long-standing argument whether EMT is a crossover or a first-order phase transition accompanied by a bistability at sufficiently low temperatures.[16-22]

In general, bistability appears in diverse systems such as electronic circuits, optoelectronic systems, magnetic systems, molecular and biological systems, etc. Figure

1(a) depicts the schematic view of bistability in the parameter space, where the function $y(x)$ exhibits a bistable region as a function of the control parameter $x$. The bistable system has one unstable blanch between two stable blanches, and the transition from one branch to another one is accompanied by a hysteresis behavior. The corresponding Ginzburg-Landau free energy is schematically shown in Fig.1(b) that has two stable equilibrium states. As the origin of bistability, a positive feedback loop existing in the system is crucial, as exemplified by the Schmitt trigger circuit, biological systems,[23-25] and the optical bistability realized with a nonlinear absorptive medium embedded in an optical cavity.[26-29]

Photoexcited e-h systems in semiconductors inherently possess such a positive feedback effect. The nature of e-h systems is determined by the thermal equilibrium condition between excitons (hydrogenlike bound e-h pairs) and e-h plasma (unbound e-h pairs) as shown in Fig. 2(a). In the low temperature and low density region the e-h system is mostly composed of excitons, while as the e-h pair density increases the screening of the long-range Coulomb interaction becomes prominent, resulting in the reduction of the binding energy of excitons $E_b$. Since unbound e-h pairs contribute to the screening of Coulomb interaction much more efficiently than the charge-neutral excitons, once some unbound e-h pairs are created among high density excitons, they facilitate the ionization of remaining excitons, which acts as the positive feedback in the EMT process. Such a positive feedback can induce a hysteresis of the ionization ratio of excitons as defined by $\alpha = n_f/n_{total}$ ($n_{total} = n_f + n_{ex}$, with $n_f$ and $n_{ex}$ the densities of free e-h pairs and excitons, respectively) with respect to the total pair e-h density, invoking the possibility that EMT becomes a first-order phase transition.[16-22] Experimentally, the problem of EMT has long been studied using e.g. photoluminescence spectroscopy.[30-33] Recent terahertz spectroscopy has shed new light on this problem,[34-37] making it possible to determine the exciton ionization ratio $\alpha$ through the quantitative evaluation of excitons and unbound e-h pairs densities. In particular, our recent study on the ultrafast dynamics of EMT in GaAs under the resonant excitation of excitons has indicated the signature of first-order phase transition.[38]

In this letter, we first show that ionization ratio of excitons exhibits a bistability as a function of total e-h pair density at low temperatures, based on a simple thermodynamic model described by the Saha equation with taking into account the empirical expression for the band-gap renormalization effect caused by the unbound e-h pairs. Furthermore, we discuss the ionization dynamics of excitons after their impulsive photoexcitation by using a rate equation analysis, and show that a peculiar incubation behavior appears on the verge of the bistable region.

First we discuss the ionization ratio of excitons, $\alpha$, at thermal equilibrium under Maxwell-Boltzmann statistics. As schematically shown in Fig. 2(a), we consider two states: the $1s$ exciton state and the unbound e-h pair state. Then, $\alpha$ is described by the Saha equation,

$$\frac{\alpha^2}{1-\alpha} = \frac{1}{n_{\text{total}} \lambda_T^3} \exp\left(\frac{-E_b}{k_B T}\right) \qquad (1)$$

where $\lambda_T = h/\sqrt{2\pi \mu k_B T}$ and $\mu = m_e m_h / m_{\text{ex}}$, with $m_e$, $m_h$ and $m_{\text{ex}}$ being the mass of electrons, holes, and excitons. The quantity $\lambda_T$ takes the form of thermal de Broglie wavelength, whereas the mass is expressed by the e-h reduced mass. Here we use $\mu = 0.044 m_0$ with $m_0$ the bare electron mass, assuming the case of bulk GaAs. We plot $\alpha$ at different temperatures in Fig. 2(b), calculated by Eq. (1) with a fixed value of $E_b = E_0$ = 4.2 meV where $E_0$ is the exciton binding energy of GaAs. $\alpha$ monotonically increases as the temperature increases reflecting the thermal ionization of excitons. Interestingly, when the pair density decreases under constant temperatures, $\alpha$ increases monotonically. This behavior reflects the mass action law, or in other words the entropy effect.[19,21,37,39] Next, to incorporate the reduction of the binding energy of excitons $E_b$ in Eq. (1) due to the screening, here we adopt the empirical expression for the band-gap renormalization derived by Vashishta and Kalia[40] and assume the energy level of $1s$ exciton to be constant regardless of the pair density, because of the charge neutrality of excitons.[41,42] The resulting curve of $E_b$ is plotted in Fig. 2(c) as a function of the e-h pair density. Although the precise description on the reduction of exciton binding energy requires more sophisticated theoretical treatment,[21] we can capture the overall behavior of the system qualitatively with such an empirical model. First, we discuss the case without a positive feedback by assuming that both excitons and unbound e-h pairs contribute to the screening equally. In this case, the horizontal axis in Fig. 2(c) should be read as the total pair density $n_{\text{total}}$. By inserting the density dependence of $E_b$ plotted in Fig.2(c) into Eq. (1), the ionization ratio $\alpha$ is calculated for different temperatures as shown in Fig. 2(d). At high temperatures most of excitons are thermally ionized, and the behavior of $\alpha$ is similar to that in Fig. 2(b). At low temperatures where excitons become dominant in the low density region, $\alpha$ exhibits a sharp upturn around the density $n_M \sim 2 \times 10^{16} \text{cm}^{-3}$ where $E_b$ becomes zero. Although the increase of $\alpha$ becomes steeper at lower temperatures, the hysteresis never appears in the $\alpha$-$n_{\text{total}}$ curve in this case. Next, we investigate the case with a positive feedback, assuming that only the unbound free e-h pairs consisting of e-h plasma contribute to the Coulomb screening. In this case, the reduction of exciton binding energy should be plotted as a function of unbound free e-h pair density $n_f$ as shown in Fig. 3(a). The calculated curves for $\alpha$ as a function of total e-h pair density are shown in Fig. 3(b). Now a hysteresis in the $\alpha$-$n_{\text{total}}$ curve appears in the low temperature region below $T_c = 5.5$ K $\sim 0.1 E_0/k_B$. Because of the self-stabilization of excitons, $\alpha$ can take a small value even above $n_{\text{total}} = n_M$. The comparison between Fig. 2(d) and 3(b) clearly shows that the positive feedback effect in the Coulomb screening is a key for the bilstability in $\alpha$. The onset temperature of bistability, $T_c$, depends on the steepness of feedback effect, which corresponds to the curvature of the reduction of $E_b$ in Fig. 3(a) in this model. As the curvature becomes steeper, $T_c$ increases and even exceeds $E_0/k_B$.

Next we discuss the dynamics how the system reaches the thermal equilibrium after the photoexcitation. For this end, we consider rate equations for the populations of excitons and unbound e-h pairs as schematically shown in Fig. 2(a),

$$dn_{\text{ex}}/dt = -\gamma_1 n_{\text{ex}} + \gamma_2 n_f^2 \quad (2)$$

$$dn_f/dt = \gamma_1 n_{\text{ex}} - \gamma_2 n_f^2 \quad (3)$$

where $\gamma_1$ represents the ionization rate of an exciton to an unbound e-h pair and $\gamma_2$ represents the formation rate of an exciton from an unbound e-h pair. In Eqs. (2) and (3), the total pair density $n_{\text{total}} = n_{\text{ex}} + n_f$ is assumed to be time-independent after the photoexcitation, as we consider here the dynamics of the e-h system in much shorter timescale than the recombination lifetime of e-h pairs. Correspondingly, the e-h system is assumed to reach the quasi-thermal equilibrium within the lifetime. At thermal equilibrium, the right hand side of Eqs. (2) and (3) can be set to zero, and from the detailed balance condition, one finds that the ratio $\gamma_1/\gamma_2 n_{\text{total}}$ is identical to the right-hand side of Saha equation (Eq. (1)), with $T$ representing the bath temperature. Then, Eqs. (2) and (3) can be rewritten in the form

$$d\alpha/dt = (\gamma_2/\lambda_T^3) \times [e^{-\beta E_b}(1-\alpha) - n_{\text{total}} \lambda_T^3 \alpha^2]. \quad (4)$$

Here we simply take $\gamma_2$ to be constant (or to depend only on temperature). In reality,

$\gamma_2$ would depend on $\alpha$, which should be treated based on a microscopic theory in a self-consistent manner. We leave this issue as a subject of future investigation. Experimentally, we can realize an initial condition $\alpha = 0$ by the resonant photoexcitation of excitons. In Fig. 4(a) we replot an $\alpha$-$n_{\text{total}}$ curve at 7 K, picked from Fig. 3(b), which is above the onset temperature of bistability. We discuss how $\alpha$ evolves into the equilibrium value from the initial condition of $\alpha = 0$ at different photoexcited e-h pair densities, as schematically shown by the arrows in Fig. 4(a). Figure 4(b) plots the temporal evolution of $\alpha$ calculated from Eq. (5) for different pair densities $n_{\text{total}}$. $\alpha$ increases monotonically in time, and approaches to the equilibrium value faster as $n_{\text{total}}$ increases.

Next, we show the behavior at 4 K where the bistability appears in the $\alpha$-$n_{\text{total}}$ curve (Fig. 5(a)). One can see that a peculiar dynamics appears on the verge of the bistable region as shown in Fig.5(b). Above the critical density as denoted by $p$ in Fig. 5(a), $\alpha$ exhibits two-step dynamics: first $\alpha$ reaches a metastable value and stays for a finite time, and then increases to a stable value $\sim 1$. As the total density $n_{\text{total}}$ approaches to the edge of the bistable region, i.e. to the saddle-node bifurcation point $p$ from the high density region, this incubation time becomes longer and shows a critical slowing down behavior as plotted in Fig. 5(c). The incubation time emerges in a wider density region at lower temperatures, where bistability appears in a wider range as shown in Fig. 3(b). Such a peculiar temporal behavior accompanied by the incubation time is consistent with the recent observation on the dynamics of EMT in GaAs,[38] indicating that EMT becomes a first-order phase transition at low temperatures. The emergence of incubation time has been observed in a variety of photoinduced phase transition phenomena, e.g. in spin-crossover complexes,[10] and discussed theoretically,[11,12] suggesting the generic character of photoinduced phenomena associated with the first-order phase transition.

In summary, we numerically investigated the dynamics of EMT. We showed that bistability emerges in the exciton ionization ratio around the EMT at sufficiently low temperatures by using Saha equation with adopting the empirical expression for the band-gap renormalization effect. Rate equation analysis shows that the EMT from the photo-created oversaturated exciton gas phase is accompanied by incubation time. A slowing down behavior appears on the density dependence of incubation time toward the bifurcation point in the exciton ionization ratio. In the present study, we performed the analysis in a classical limit where e-h systems obey the Maxwell-Boltzmann statistics and the Saha equation is validated. It is a highly intriguing issue to extend this work to the quantum degenerate regime where exciton Bose-Einstein condensation or e-h BCS state emerges, by incorporating the quantum effect and to study the non-equilibrium dynamics of these quantum phases.

**Acknowledgemnts**

This work was supported by JSPS KAKENHI Grant Numbers 22244036 and 15H02102.

**Figures and Figure captions**

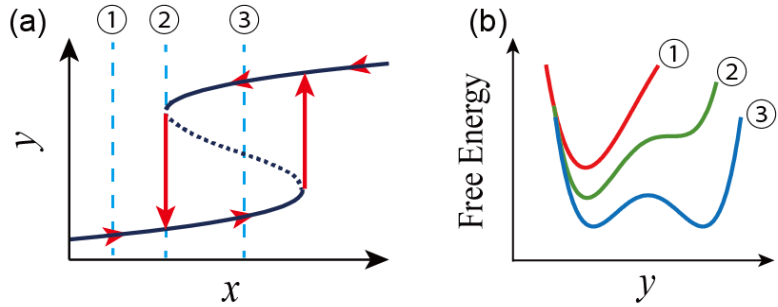

Fig. 1 Schematic picture of a bistable system. (a) Output $y$ as a function of input $x$. (b) Free energy at each cut in Fig. 1(a).

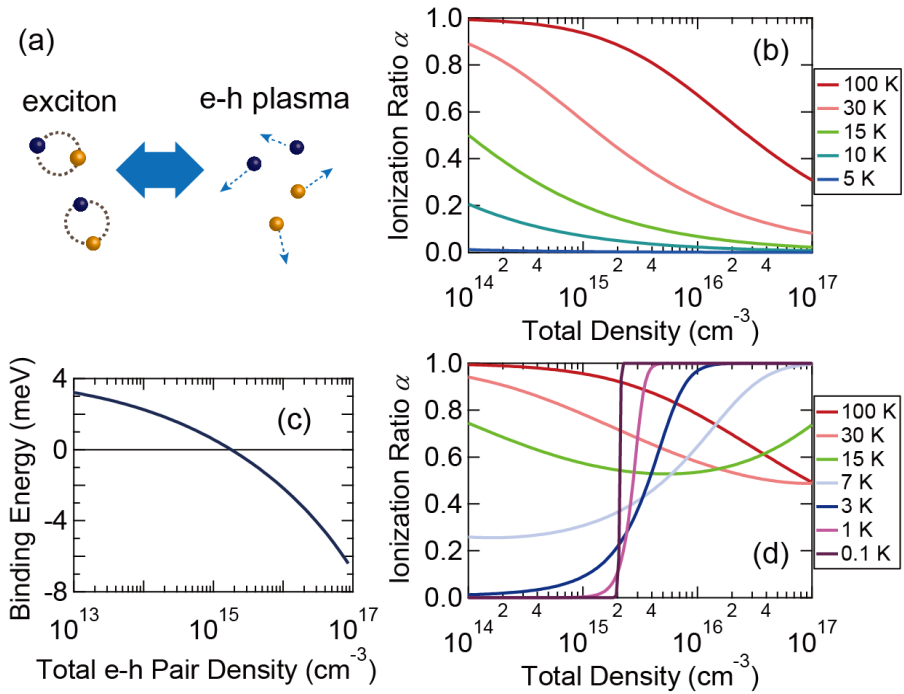

Fig. 2 (a) Schematic picture of exciton ionization into unbound e-h plasma and exciton formation from e-h plasma. (b) Ionization ratio of excitons, $\alpha$, at thermal equilibrium at different temperatures, calculated with a fixed binding energy $E_b = 4.2$ meV assuming the case of bulk GaAs. (c) The exciton binding energy $E_b$ calculated using the expression of band-gap renormalization by Vashishta and Kalia.[40)] Horizontal axis corresponds to the total pair density $n_{total}$. (d) Ionization ratio of excitons $\alpha$ at thermal equilibrium at different temperatures, calculated with a density-dependent binding energy $E_b$ shown in Fig. 2(c).

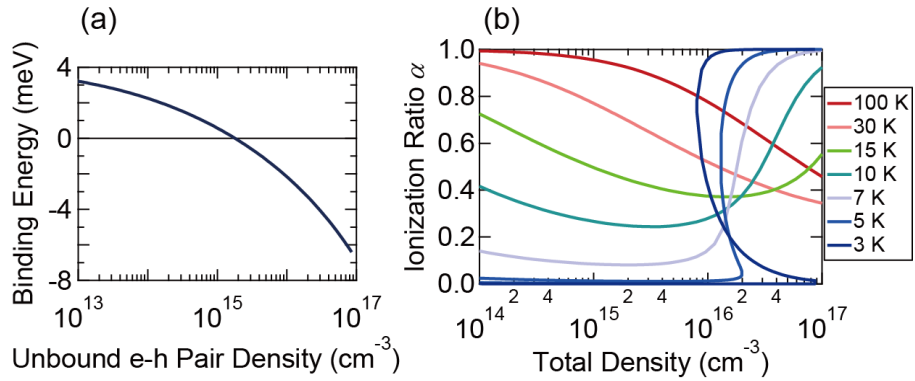

Fig. 3 (a) The density dependence of exciton binding energy, $E_b$. Horizontal axis corresponds to the unbound free e-h pair density, $n_f$. (b) Ionization ratio of excitons $\alpha$ at thermal equilibrium at different temperatures, calculated with $E_b$ shown in Fig. 3(a).

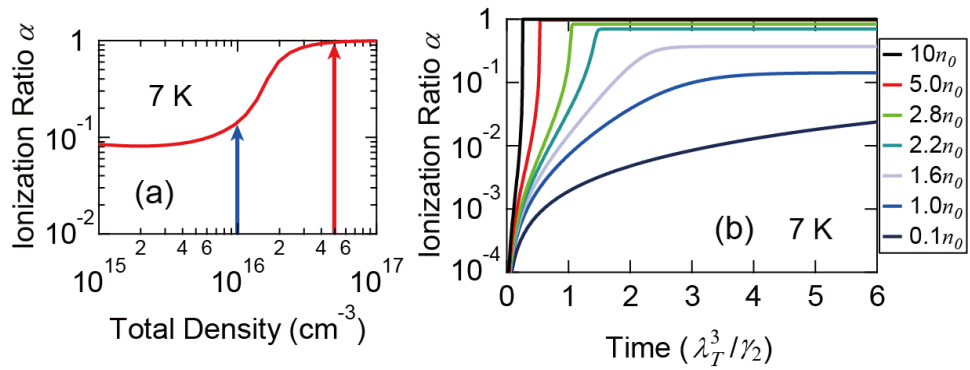

Fig. 4 Dynamics of exciton ionization after the resonant photoexcitation of excitons, i.e. starting from $\alpha = 0$. (a) $\alpha$ at thermal equilibrium at $T = 7$ K $> T_c$. Arrows indicate the paths $\alpha$ follows. (b) Temporal evolution of $\alpha$ at the indicated total e-h pair densities where $n_0 = 1 \times 10^{16}\,\mathrm{cm}^{-3}$.

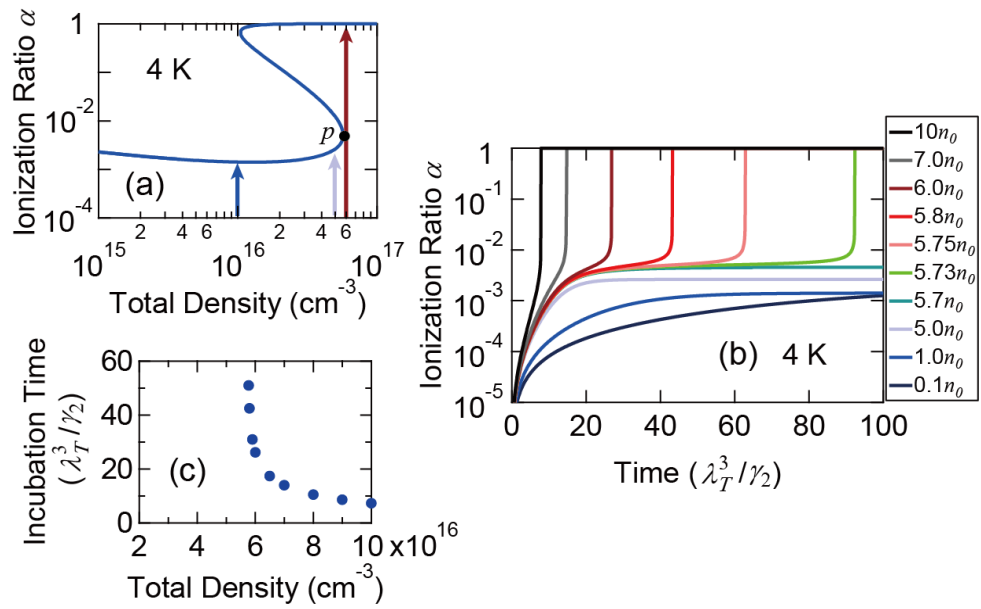

Fig. 5 Dynamics of exciton ionization after the resonant photoexcitation of excitons, i.e. starting from $\alpha = 0$. (a) $\alpha$ at thermal equilibrium at $T = 4$ K $< T_c$. Arrows indicate the paths $\alpha$ follows. (b) Temporal evolution of $\alpha$ at different photoexcited total densities, where $n_0 = 1 \times 10^{16}$ cm$^{-3}$. (c) Density dependence of the incubation time.